# Hydrogen in the LCLS2 Beamline Vacuum

Anthony C. Crawford     Fermilab Technical Div./SRF Development Dept.     acc52@fnal.gov     13Oct15

This note demonstrates that the cold segments of the LCLS2 linac will cryopump hydrogen for long periods of time, mitigating concerns for beam-residual gas interactions.

## Introduction

Stainless steel desorbs hydrogen when it is at room temperature unless special care is taken to remove dissolved hydrogen from the bulk of the steel. The customary way to degas stainless steel is at a temperature above 900C in ultrahigh vacuum. The rate of hydrogen desorption from room temperature stainless steel surfaces in a linac will be large enough to make hydrogen the dominant gas species in the vacuum if adsorbed water molecules have been removed from the surfaces. Superconducting RF cavity surfaces operating at 2K will act as a cryopump for the hydrogen molecules.

It is known from experience at DESY that for T = 4.2K, superconducting niobium cavities will stop cryopumping hydrogen after the accumulation of approximately one monolayer on the cavity surface [1], resulting in a large accumulation of hydrogen gas in cold regions. What is the case for SLAC Linac Coherent Light Source-II (LCLS2) cryomodules operating at 2K? Is it necessary to hydrogen degas stainless steel components in warm regions? In order to investigate these questions we do the following things:

1. Review the literature for cryopumping of hydrogen on metal surfaces
2. Review measurements from the recent test of Fermilab cryomodule number 2 (CM2)
3. Review the estimate for hydrogen accumulation in the Tesla Test Facility (TTF) capture cavity
4. Update and refine the TTF estimate using CM2 measurements

## Literature Review: Hydrogen on Cold Surfaces

The ability of a cold surface to cryopump hydrogen is a balance between the processes of condensation and desorption, or, expressed in the terminology of vacuum physics, between the "sticking coefficient" and the desorption rate for hydrogen molecules. The sticking coefficient is the probability that any hydrogen molecule will be condensed on the cold surface and varies between zero and one. A sticking coefficient of one means that every molecule will be condensed. Measured values for sticking coefficient and desorption rates are shown in Figure 1 through Figure 3. These figures and their captions are from reference [2].

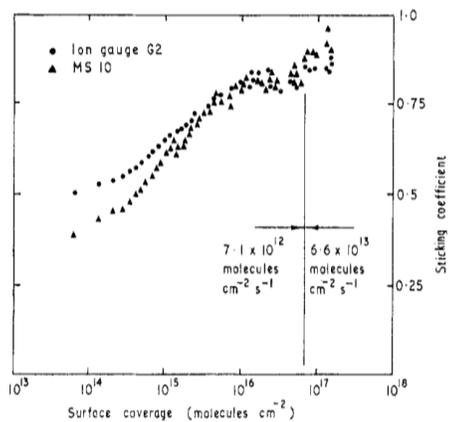

Figure 1. Example of variation of sticking coefficient with surface coverage for hydrogen at 300°K incident on to surface at 3·14°K (3.05.66).

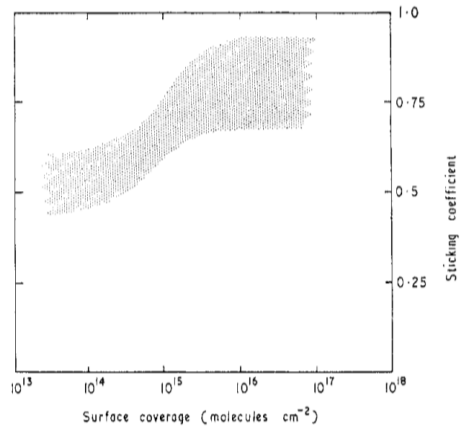

Figure 2. General variation of sticking coefficient with surface coverage for hydrogen at 300°K incident on to surfaces ranging from 2·17 to 3·68°K and with incidence rates from $8 \times 10^{11}$ to $2 \times 10^{13}$ molecules $cm^{-2}$ $s^{-1}$.

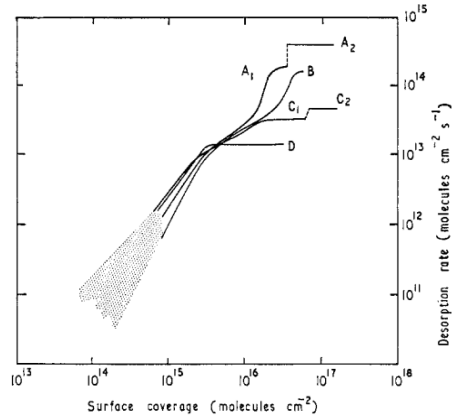

Figure 3. Variation of desorption rate with surface coverage for hydrogen at 300°K condensed on to surfaces at different surface temperatures. $A_1 A_2$, 3·68°K at $5·6 \times 10^{12}$ and $3·5 \times 10^{13}$ molecules cm$^{-2}$ s$^{-1}$ respectively. B, 3·37°K at $9·2 \times 10^{12}$ molecules cm$^{-2}$ s$^{-1}$. $C_1$, $C_2$, 3·14°K at $7·1 \times 10^{12}$ and $6·6 \times 10^{13}$ molecules cm$^{-2}$ s$^{-1}$ respectively. D, 2·18°K at $1·2 \times 10^{13}$ molecules cm$^{-2}$ s$^{-1}$.

It is the 2.17K data in the figures that interests us most since it is closest to the operating temperature of the cavities for LCLS2. For a molecule the size of hydrogen, one monolayer corresponds to ~3 x $10^{15}$ molecules/cm$^2$ [2]. It is evident from Figure 2 that the sticking coefficient increases by less than a factor of two for increasing surface coverage up to one monolayer, while from Figure 3, we see that the desorption rate increases by orders of magnitude. This is why effective cryopumping stops at 4.2K for thicknesses greater than one monolayer. The desorption rate grows rapidly for increasing thickness. However, the desorption curves plateau at lower and lower values as surface temperature is decreased. At 2K the desorption rate is anticipated to plateau at a smaller value than curve D in Figure 3. This should mean that a 2K surface should have a desorption rate approximately two orders of magnitude lower than for a 4.2K surface.

Measurements for Figure 1 through 3 were made with hydrogen inflow rates greater than 5 x $10^{12}$ molecules/cm$^2$-sec. The expected inflow rate for a single nine cell 1.3 GHz cavity with a warm beam tube at each end is 1 x $10^9$ molecules/cm$^2$-sec [3]. According to reference [2] the accumulation rate has a major effect on the surface condition of the condensate, in particular, the size of hydrogen crystals, which in turn affects the sticking coefficient. The result is, for T = 3.68K, shown in Figure 4. Figure 4 and its caption are sourced from reference [2].

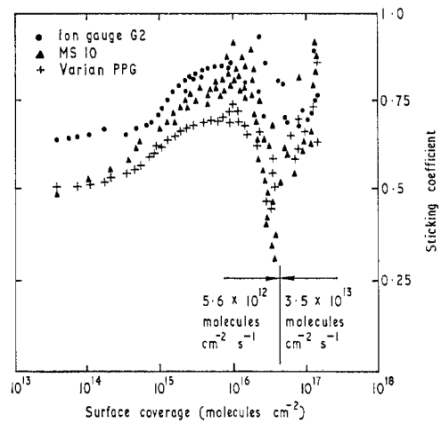

Figure 4. Example of variation of sticking coefficient with surface coverage for hydrogen at 300°K condensing on to surface at 3·68°K (28.07.66).

For slow accumulation rates, less than $5 \times 10^{12}$ molecules/cm$^2$-sec, the sticking coefficient decreases rapidly after a few monolayers at 3.68K. What is the case for 2K and $1 \times 10^9$ molecules/cm$^2$-sec? Data for this case has not been found in the literature with the exception that reference [4] states, in disagreement with reference [2], that no dependence on the rate of deposition was seen for rates as low as $7 \times 10^{10}$ molecules/cm$^2$-sec. Thus, determination from published results of the sticking coefficient for hydrogen at 2K after one monolayer of slow accumulation is problematic.

**Measurements**

Fortunately, we have data for a real cryomodule. The temperature and pressure history of a recent test of FNAL CM2 is shown in Figure 5. The vertical axis legend is as follows:

N:M1CVSM   =   integrated cavity accelerating voltage for the cryomodule  [MV]
N:M1CCU    =   upstream end beamline pressure from cold cathode gage  [Torr]
N:M1IPU    =   upstream end beamline pressure from ion pump  [Torr]
N:M1CCD    =   downstream end beamline pressure from cold cathode gage  [Torr]
N:M1IPD    =   downstream end beamline pressure from ion pump  [Torr]
N:3TX03    =   helium temperature  [K]

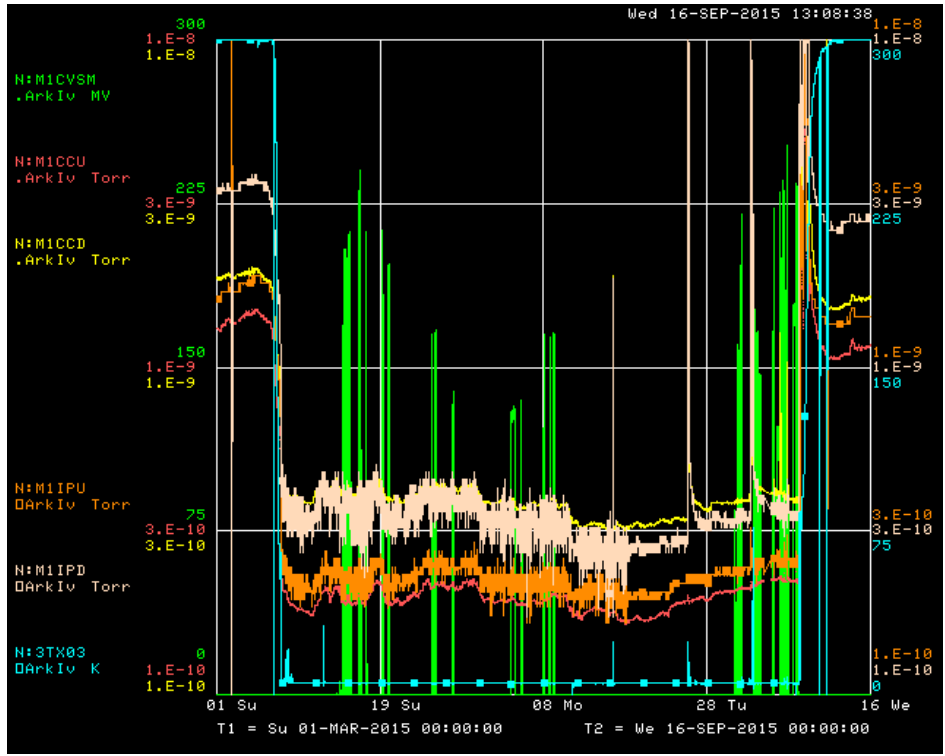

Figure 5.  Fermilab CM2 Temperature and Pressure
(Data provided by E. Harms of FNAL/AD)

There is no apparent reduction in cryopumping for the ~ 150 day duration at T = 2K. The pressure in the warm beamtubes at the ends of the cryomodule ranges from $P = 2 \times 10^{-10}$ Torr to $P = 4 \times 10^{-10}$ Torr. Ion pumps at the ends of the cryomodule are responsible for some of the pumping speed for hydrogen.

The assumptions for gas load and pumping speed in the Tesla Test Facility (TTF) estimate of reference [3] agree substantially with the observed pressure readings at both ends of FNAL CM2. An infinite capacity for cryopumping hydrogen and a 70mm cavity iris diameter conductance limitation was assumed for the TTF estimate. Table 1 compares the TTF estimate to measured values at CM2. Symbols and formulas from TTF have been used for ease of comparison. Units of Torr for pressure, rather than milliBar as per TTF, have been adopted. The expression for the number of molecules entering the cold region per unit time is the time derivative of the ideal gas law:

$$\frac{d}{dt}(PV) = \dot{Q} = \dot{n}kT,$$

as per reference [3].

| Parameter | Symbol | Mathematical Expression | Units | TTF Estimate | CM2 Measurement | CM2 Estimate |
|---|---|---|---|---|---|---|
| Outgassing rate of hydrogen from non-degassed stainless steel | | | Torr-liter/sec-cm$^2$ | 7.5 x 10$^{-9}$ | | |
| Surface area of stainless steel | $F$ | | cm$^2$ | 7π(200) | | |
| Cold surface area | | | cm$^2$ | 1 x 10$^4$ | | 8 x 10$^4$ |
| Total hydrogen gas flow to cold region | $\dot{Q}$ | $\frac{d\dot{Q}}{dF} \cdot F$ | Torr-liters/second | 3.3 x 10$^{-7}$ | | 1.8 x 10$^{-7}$ |
| Pumping Speed of cold region | $S_{eff}$ | | liters/second | 600 | | |
| Gas pressure at entrance to cold region | $P_{ext}$ | | Torr | 5.3 x 10$^{-10}$ | 3 x 10$^{-10}$ | |
| Number of molecules entering the cold region per unit time | $\dot{n}$ | $\dot{Q} \cdot \frac{1}{kT}$ | molecules/second | 1 x 10$^{13}$ | | 5.8 x 10$^{12}$ |
| Boltzmann Constant | $k$ | 1.04 x 10$^{-22}$ | Torr·liters/K | | | |

Table 1    TTF and CM2 Parameters

The methodology used to compare the TTF and CM2 values is as follows:

We wish to estimate the number of hydrogen molecules entering the cold region per unit time for CM2 by using the measured value for pressure at each end of the cryomodule rather than relying on the estimated value for TTF. For this we use:

$$\dot{Q} = P_{ext} \cdot S_{eff} = 3\text{x}10^{-10} \text{ Torr} \cdot 600 \text{ liters/second} = 1.8\text{x}10^{-7} \text{ Torr·liters/second}$$

The number of hydrogen molecules entering the cold region, based on CM2 pressure measurement is then given by:

$$\dot{n} = \dot{Q} \cdot \frac{1}{kT} = 1.8\text{x}10^{-7} \text{ Torr-liters/second} \cdot (1.04 \text{ x } 10^{-22} \text{ Torr·liters/K} \cdot 300\text{K})^{-1} = 5.8 \text{ x } 10^{12} \text{ molecules/second}$$

Using the value of 3 x 10$^{15}$ H$_2$/cm$^2$ for one monolayer of coverage from reference [2], 5.8 x 10$^{12}$ H$_2$/second entering the cold region of CM2, and 8 x 10$^4$ cm$^2$ for the cold surface area of a cryomodule containing eight cavities results the following accumulation rate for CM2:

$$\frac{5.8 \text{ x } 10^{12} \text{ H}_2/\text{second}}{8 \text{ x } 10^4 \text{ cm}^2} = 7.3 \text{ x } 10^7 \frac{H_2}{sec \cdot cm^2} = 7.3 \text{ x } 10^7 \frac{H_2}{sec \cdot cm^2} \cdot \frac{1 \text{ } cm^2}{3\text{x}10^{15} \text{ H}_2} = 2.4 \text{ x } 10^{-8} \text{ monolayers/second}$$

$$2.4 \times 10^{-8} \text{ monolayers/second} = 0.0021 \text{ monolayers/day} = 0.77 \text{ monolayers/ year}$$

The value used here for one monolayer of coverage is a factor of three larger than was used for the TTF estimate. This is a significant difference and means that CM2 had accumulated only ~ 0.3 monolayers of hydrogen, averaged over the whole cold surface, at the end of the 150 day cold test. The estimated accumulation time for one monolayer of hydrogen for a single cryomodule containing eight 1.3 GHz cavities, is 1.3 years. As an additional example, this same methodology is applied to earlier measurements from DESY/HERA cryomodules, circa 1993. The calculations and results are exhibited in the second part of the appendix to this note.

There is no consistent correlation between pressure and electric field in the cavities shown in Figure 5. The excursions to higher pressure at the downstream end of the cryomodule during the final weeks of the test happen in one instance when the cavities are not powered. This is an important result and means that there is no noticeable effect on the rate of desorption by field emitted electrons, field emission induced x-rays, or RF electric fields. An integrated cryomodule voltage equal to 225 MV is equivalent to an accelerating gradient of 28 MV/m in the cavities. This is a much higher gradient than the LCLS2 operating specification of 16 MV/m. Even though this result is reassuring, it is not possible to be certain that it will carry over to LCLS2 cryomodules because the RF duty factor for CM2 is more than a factor of 100 lower than for the continuous wave (CW) operation of LCLS2. Final results must wait for the first CW cryomodule test in 2016.

It should be noted that the stainless steel beamtubes and other components for the CM2 test were not hydrogen degassed. Each end of the cryomodule was pumped by an ion pump with a nominal pumping speed of 120 liters/minute for nitrogen and ~ 250 Liters/minute for hydrogen. This means that approximately half the hydrogen degassing in the warm regions was pumped by the ion pump, with the rest entering the cold region.

### Implications of Hydrogen Accumulation for Cavity Performance

Hydrogen accumulation in the cryomodules is good for the particle beam. It means that there are fewer molecules available to interact with the beam. For LCLS2, the presence of a one microsecond time interval between populated RF buckets means that molecules ionized by the electron bunch will disperse before the next trailing bunch arrives and that a fast ion instability will not happen [5]. Lower residual gas density means lower probability for single electron bunch effects. Beam-gas scattering effects are reduced as well.

However, accumulation of condensates is potentially problematic for cavity performance, with reported occurrences of both increased field emission and increased RF surface resistance after the accumulation of one monolayer of gas [6]. Pulsed power processing and RF helium discharge conditioning have been used in the past to correct this condition, with some success. Reference [7] records an instance of increase in the residual resistance of a cavity by a factor of three due to cryopumped gas. This is a particularly interesting example because the cavity being tested was free of field emission. The cavity performance is shown in Figure 6, reproduced from reference [7].

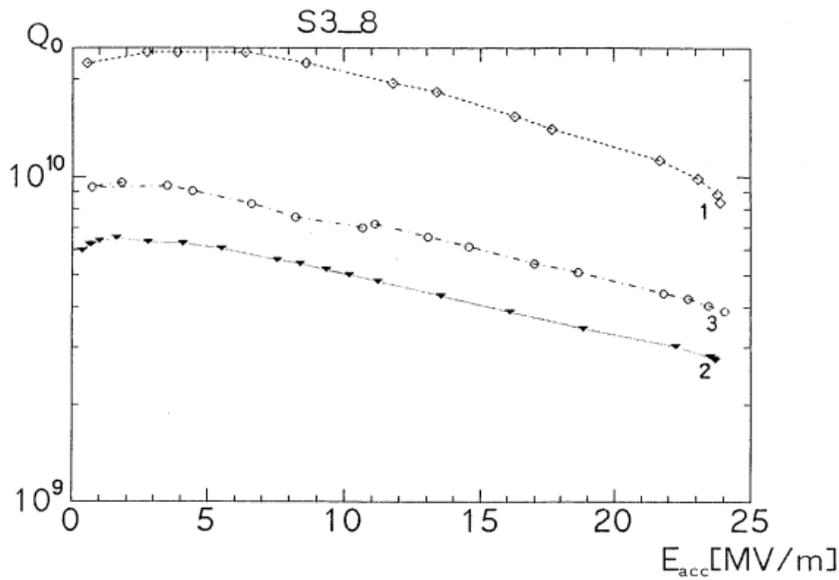

Fig. 6 Change of the residual surface resistance of a 3-GHz single cell cavity due to adsorption of residual gases. (1): initial performance, (2): with cryopumped gases, initially frozen in the pumping tube. (3): after cryocycle to 300 K

If surface resistance increase due to hydrogen accumulation were to happen in a cryomodule, the effect would start with the cavity cells at each end of the module and progress toward the center of the module with time. It will be interesting to watch for this in the future for continuous cold operation approaching one year in duration.

Better accuracy is obtained in measurement of cavity surface resistance when operation is in the continuous wave mode. This is because the higher duty factor increases the ratio of RF power loss to the static heat load in the cryomodule. Measurements of RF loss are usually made by estimating the rate at which liquid helium changes to the gas phase when the cavities are in operation. LCLS2, because it operates in the continuous RF power mode and because of the extremely low surface resistance of its nitrogen doped cavities, should provide an excellent opportunity to observe any cavity surface resistance degradation due to long periods of time at 2K.

### Hydrogen Accumulation for LCLS2

Table 2 lists the estimated time for accumulation of one monolayer of hydrogen for each of the linac segments in LCLS2, assuming performance equal to CM2.

| Linac Segment | Number of Cavities | Time to Accumulate One Monolayer [days] |
|---|---|---|
| L0 | 8 | 475 |
| L1 | 16 | 950 |
| HL | 16 | 950 |
| L2 | 88 | 5225 |
| L3 | 160 | 9500 |

Table 2  Time to Accumulate One Monolayer for LCLS2 Linac Segments

It is anticipated that the linac segments will be brought to room temperature at least once per year. This will allow removal of any accumulated hydrogen from the cold sections.

**Implications of 2K Operation for SRF in Accelerators and Storage Rings**

The following important advantages accompany operation below 2.17K and should be taken in to consideration when choosing the operating temperature for future SRF systems:

- Increased cryopumping capacity for hydrogen (mitigates ion trapping and beam instability issues for the case of long trains of filled RF bunches)
- Decreased fluid noise level (mitigates problems with microphonics)

These advantages should be carefully compared with the cost savings associated with higher temperature operation, for example, when the use of high critical temperature cavities, such as $Nb_3Sn$, is contemplated.

**Conclusion**

Measurements from Fermilab CM2 indicate that there is no apparent reduction in cryopumping speed for hydrogen for an operating duration at 2K approaching one half year. The measured value for pressure at the entrance to CM2 allows us to predict a time to accumulate one monolayer of hydrogen equal to one and one third years for a cryomodule containing eight TESLA style 1.3 GHz cavities. Hydrogen degassing of warm beamline components should not be necessary for LCLS2 linac segments.

**Acknowledgements**

The author is thankful for thoughtful discussions provided by Sam Posen (Fermilab) and Wolf-Dietrich Moeller (DESY).

# Appendix

## Appendix Part 1    A Concise Cryogenic Surface Dictionary

*Absorption*
Assimilation of a molecular species throughout the bulk of a solid or liquid is termed as absorption.  Absorption of hydrogen does not take place in niobium surfaces at cryogenic temperatures.  Hydrogen uptake is prevented by a naturally occurring 2 to 3 nanometer thick coating of niobium pentoxide, $Nb_2O_5$, at the niobium surface.  In this sense, when we speak of cryopumping on to a niobium surface, what is really meant is on to an $Nb_2O_5$ surface .

*Adsorption*
Accumulation of a molecular species at the surface rather than in the bulk of the solid or liquid is termed as adsorption.  This is what happens to hydrogen impinging on niobium at 2K.

*Chemisorption*
Chemisorption is adsorption in which the forces involved are valence forces of the same kind as those operating in the formation of chemical compounds. [A1]  Chemisorption is the mechanism of titanium sublimation pumps, titanium and titanium-tantalum sputter ion pumps, non-evaporable getter pumps, and etc.

*Cryocondensation*
With increasing coverage of adsorbed gas van der Waals forces acting between the adsorbed molecules becomes an important effect.  When coverage reaches the point where the intermolecular forces dominate, the surface is said to be condensing the adsorbed gas.  During cryocondensation, the phase of the gas changes to either the liquid or solid phase.  Vacuum pumps that operate in this regime are called cryocondensation pumps.

*Cryosorption*
Vacuum pumps operating on the principle of physisorption are referred to as "cryosorption" pumps.  Because of the limited amount of surface coverage before the onset of cryocondensation, cryosorption pumps rely on very large surface areas.  Large surface area is typically achieved with the use of highly porous material such as synthetic zeolite.

*Cryotrapping*
Cryotrapping happens when a condensable gas is used to trap a non-condensable gas.   The non-condensable gas becomes incorporated in the condensed matrix of the carrier gas rather than being physisorbed on to the base substrate.  Argon can be used to trap hydrogen in cryopumps. [A3]

*Degassing*
Degassing is the deliberate removal of a gas from a solid or liquid.

*Desorption*
Desorption is a phenomenon whereby a substance is released from or through a surface.

*Outgassing*
Outgassing is the spontaneous release of gas from a solid or liquid.

*Physisorption*
Physisorption is adsorption in which the forces involved are intermolecular forces (van der Waals forces) of the same kind as those responsible for the imperfection of real gases and the condensation of vapors, and which do not involve a significant change in the electronic orbital patterns of the species involved. [A1]  For sub-monolayer coverage, physisorption is the dominant mode, with van der Waals forces acting between the adsorbed molecules and the surface.  Vacuum pumps operating on the principle of physisorption are referred to as "cryosorption" pumps. [A2]

*Sorption*
Sorption is a physical or chemical process by which one substance becomes attached to another.  The term sorption is used to describe absorption, adsorption or ion exchange.

## Appendix Part 2    DESY/HERA Cryomodules

One half of a HERA cryomodule is shown in Figure A1, sourced from reference [A4].

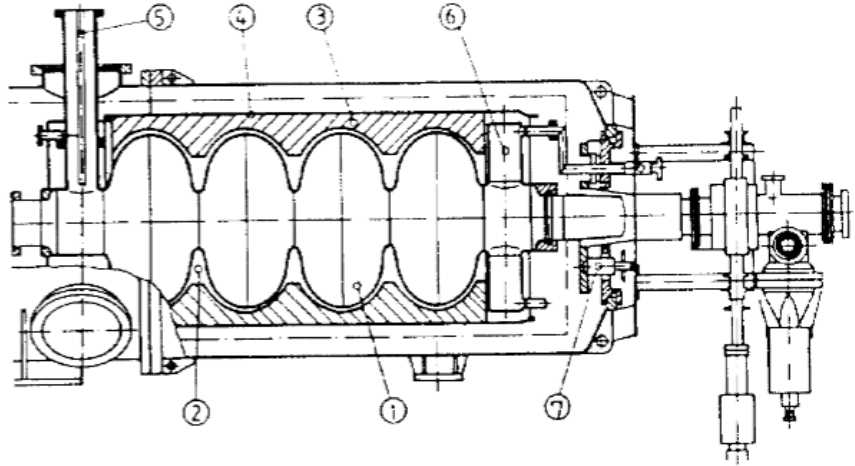

Fig.A1 Main layout of the cryostat. The total length is 5.2 m. One half is shown here.

We make the following assumptions:
1. Hydrogen desorption occurs at copper plated stainless steel pipes at each end of the cryomodule
2. The isolation valves are closed
3. The warm part of the pipes has a length of l=25 centimeters between the cavity and the isolation valve at each end of the cryomodule
4. The pipes have an average effective diameter of d=13.6 centimeters

Results are listed in Table A1.

| Parameter | Symbol | Mathematical Expression | Units | HERA Measurement | HERA Estimate (2 cavities) |
|---|---|---|---|---|---|
| Cold surface area | $A$ | | cm$^2$ | | 3.8 x 10$^4$ |
| Total hydrogen gas flow to cold region | $\dot{Q}$ | $P_{ext} \cdot S_{eff}$ | Torr-liters/second | | 6.8 x 10$^{-7}$ |
| Pumping Speed of cold region | $S_{eff}$ | $12.1 \cdot \frac{d^3}{l} \sqrt{\frac{28}{2}}$ | liters/second | | 9100 |
| Gas pressure at entrance to cold region | $P_{ext}$ | | Torr | 7.5 x 10$^{-11}$ | |
| Number of molecules entering the cold region per unit time | $\dot{n}$ | $\dot{Q} \cdot \frac{1}{kT}$ | H$_2$/second | | 2.2 x 10$^{13}$ |
| Boltzmann Constant | $k$ | 1.04 x 10$^{-22}$ | Torr·liters/K | | |
| 1 Monolayer of Hydrogen | | 3 x 10$^{15}$ | H$_2$/cm$^2$ | | |

Table A1.    HERA Cryomodule Parameters

The monolayer accumulation rate for the HERA cryomodule is then given by:

$$rate = \frac{\dot{n}}{A} \cdot \frac{1\ cm^2}{3x10^{15}\ H_2} = 1.9x10^{-7} \frac{monolayers}{sec}$$

This number is equivalent to a time to one monolayer in 60 days. This length of time is indicated in red in Figure A2 from reference [A5]. The reader should keep in mind that the monolayer accumulation rate calculated here is based on the assumption that accumulation happens uniformly over the cold surface area of the cryomodule. This is not really the case for HERA cryomodules where local cryopumping stops after a monolayer is reached in each particular part of the surface.

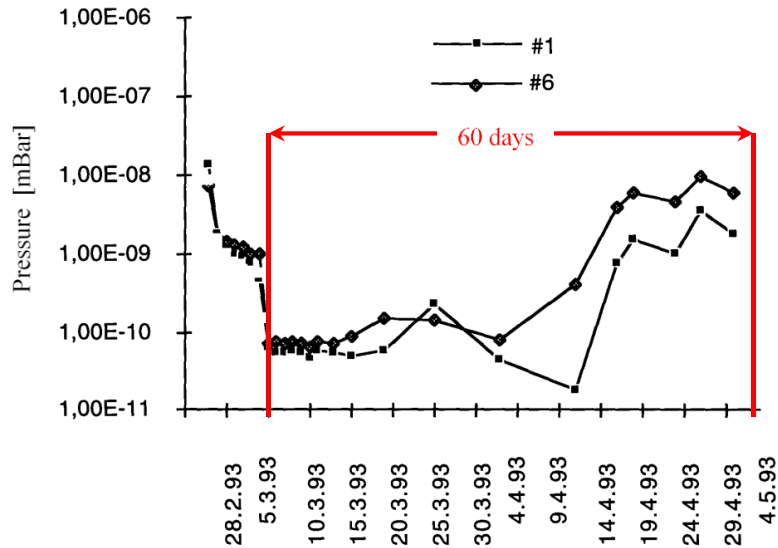

Figure A2    Cavity vacuum during and after cooldown Both gate valves in the beam line are closed. The deterioration of the beam vacuum after cooldown is due to desorbed Hydrogen

It appears in Figure A2, that the monolayer starts to form in the end cells of the cryomodule cavities after approximately 40 days and is fully formed throughout the cells of both cavities after approximately 60 days. Pumping stops after one monolayer of accumulation because of the high desorption rate at the operating temperature of 4.2K.

## Appendix References